\documentclass[%
 reprint,
 amsmath,amssymb,
 aps,
]{revtex4-2}

\usepackage{graphicx}
\usepackage{dcolumn}
\usepackage{bm}
\usepackage{xcolor} 

\usepackage[mathscr]{euscript}
\begin{document}

\preprint{APS/123-QED}

\title{Decomposition of Anomalous Diffusion \\ in two-state random walks}

\author{Abhijit Bera}
\affiliation{%
 Department of Physics, University of Houston
}%
\affiliation{
 Texas Center for Superconductivity, University of Houston
}
\author{Kevin E. Bassler}%
\email{bassler@uh.edu}
\affiliation{%
 Department of Physics, University of Houston
}%
\affiliation{
 Texas Center for Superconductivity, University of Houston
}
\affiliation{
 Department of Mathematics, University of Houston
}


\date{\today}

\begin{abstract}

Two-state stochastic models, where motion alternates between distinct dynamical modes, are widely observed in complex systems. Here we study the Two-State Random Walk (TSRW), which switches between a continuous-time random walk (CTRW) rest state and a standard Lévy walk (LW) motion state, each with power-law–distributed sojourn times. Using anomalous diffusion decomposition, we show that TSRWs exhibit a generic coexistence of Joseph (correlation), Noah (heavy-tailed increments), and Moses (aging) effects. Strikingly, although classical Lévy walks alone possess only the Joseph effect, both Noah and Moses effects emerge in TSRWs solely due to stochastic switching with the CTRW phase. Our results demonstrate that coupling between dynamical states can fundamentally reshape the mechanisms driving anomalous diffusion, offering a minimal yet powerful framework for transport in heterogeneous and intermittently switching environments.

\end{abstract}

\maketitle

\section{Introduction}

Many complex systems exhibit motion that alternates between long ballistic excursions and episodes of immobilization or localized wandering. 
Examples range from intracellular transport with intermittent binding~\cite{sabri2020elucidating,weigel2011ergodic} to animal foraging strategies involving fast relocation and slow search phases~\cite{sims2008scaling}. 
Such intermittency is a key hallmark of anomalous diffusion---where the mean squared displacement (MSD) scales as $\langle x^2(t) \rangle \propto t^{2H}$ with $H \neq 1/2$. 
Understanding the physical origin of this scaling has become increasingly urgent due to the rapid developments in single-particle tracking~\cite{shen2017single}.

A Two-State Random Walk (TSRW) model, introduced in~\cite{liu2022strong}, provides a minimal framework for describing such intermittent behavior. 
In this process, a particle stochastically switches between two dynamical modes: a continuous-time random walk (CTRW) rest state characterized by waiting periods with zero velocity, and a Lévy-walk (LW) motion state in which the particle moves ballistically with constant speed. 
Both sojourn times follow power-law distributions, giving rise to strong nonstationarity and memory effects. A similar type of two-state model~\cite{PhysRevLett.71.3975,WEEKS1996291,SOLOMON199470} was introduced to describe the transport of tracer particles in laminar fluid flow within a rotating annulus.

While this model is simple in construction, the origin of its anomalous diffusion is surprisingly subtle: 
the LW state alone possesses only temporal correlations (Joseph effect), without heavy-tailed step-size fluctuations or aging; 
conversely, the CTRW state produces aging and broad waiting-time statistics, but lacks persistent increments. 
Only through their stochastic coexistence do all three mechanisms of anomalous diffusion emerge simultaneously.

To explain this interplay, we adopt the anomalous diffusion decomposition framework developed in~\cite{chen2017anomalous,aghion2021moses}. 
This formalism attributes deviations from normal diffusion to three constitutive effects: 
(i) the Joseph effect---long-range increment correlations, 
(ii) the Noah effect---heavy-tailed increment distributions, and 
(iii) the Moses effect---temporal nonstationarity or aging. 
Each effect is associated with an exponent ($J, L, M$), and together they satisfy the universal scaling relation
\begin{equation}
H = J + L + M - 1,
\label{eq:scaling_intro}
\end{equation}
linking microscopic mechanisms to macroscopic diffusion.

Since its introduction, the decomposition framework has been generalized to a wide range of stochastic and deterministic systems. It was extended to aging deterministic dynamics~\cite{meyer2018anomalous}, applied to Generalized Lévy Walks (GLWs) in the big-jump regime~\cite{aghion2021moses}, and more recently to Variable-Speed Generalized Lévy Walks (VGLWs)~\cite{jzwq-xymr}. Beyond theoretical studies, the framework has been used to analyze diverse empirical systems, ranging from molecular transport and animal migration~\cite{vilk2022unravelling} to biological and financial time series~\cite{trillot2025evidence,meyer2023return,barraza2025non,zamani2021anomalous,salek2024statistical,salek2024equity}. In parallel, machine-learning approaches have been developed to infer diffusion models and estimate the corresponding scaling exponents directly from trajectory data~\cite{meyer2022decomposing,munoz2021objective,argun2021classification,garibo2021efficient,malinowski2025cinnamon}.

Stepping back, classical normal diffusion results from the Central Limit Theorem (CLT): 
when increments are independent, identically distributed, and with finite variance, the displacement distribution becomes Gaussian and $\langle x^2(t)\rangle \sim t$~\cite{laplace1810approximations,fischer2011history}. 
Brownian motion is the archetype of such behavior~\cite{brown2022miscellaneous}. 
However, numerous experiments---from blinking quantum dots~\cite{plakhotnik2010anomalous,stefani2009beyond} to cold atoms in optical lattices~\cite{dechant2012anomalous, RevModPhys.95.031003} and financial markets~\cite{bassler2007nonstationary,chen2017anomalous}---violate CLT assumptions, revealing anomalous diffusion that requires a more nuanced interpretation.

Traditional models such as CTRWs, Lévy flights and Lévy walks~\cite{montroll1965random,shlesinger1982random,shlesinger1993strange}, and Generalized Lévy Walks (GLWs)~\cite{shlesinger1987levy} explain many such cases, but fail to account for the combined effects produced by intermittent switching. 
Hybrid models incorporating multiple mechanisms---e.g., CTRW--FBM mixtures~\cite{fox2021aging} and two-state Langevin processes~\cite{miyaguchi2016langevin}---demonstrate that anomalous diffusion is often emergent rather than tied to a single underlying law.

In this work, we show that the TSRW provides a particularly clean representation of this emergent behavior. 
Despite consisting of two well-understood components, neither of which alone exhibits all three constitutive effects, their coupling gives rise to rich anomalous dynamics. 
We map out the full parameter space and identify four dynamical phases, each dominated by different combinations of Joseph, Noah, and Moses contributions. 
The decomposition not only clarifies the origin of anomalous diffusion in TSRWs but also provides theoretical guidance for interpreting intermittently driven transport observed in real systems.




\section{General Description of the Model}

We consider a one-dimensional stochastic process in which a particle alternates between a continuous-time random walk (CTRW) phase and a Lévy walk (LW) phase~\cite{liu2022strong,liu2022coexistence}. 
The dynamics begin with a resting period of random duration $\tau_r$, drawn from a waiting-time distribution $\omega_r(\tau)$ with a power-law tail,
$\omega_r(\tau) \sim \tau^{-\alpha}$.
During this CTRW phase, the particle remains immobile. 
At the end of the resting period, the particle performs an instantaneous spatial jump of length $\Delta x_r$, sampled from a Gaussian jump-length distribution $\lambda(x)$.

Following the CTRW phase, the particle enters the Lévy walk phase, during which it moves ballistically at a constant speed $v$ in a randomly chosen direction, either forward or backward with equal probability.
The duration of this ballistic motion, denoted $\tau_j$, is drawn from a second waiting-time distribution $\omega_j(\tau)$, which also exhibits a power-law decay,
$\omega_j(\tau) \sim \tau^{-\beta}$.
The displacement during the LW phase is therefore $\Delta x_j = \pm v \tau_j$.
This ballistic motion can be described by the coupled space--time probability density function
\begin{equation}
\phi(x,t) = \frac{1}{2}\,\delta(|x|-vt)\,\omega_j(t).
\end{equation}

Upon completion of the LW phase, the particle returns to the resting state and a new CTRW phase begins.
This alternating sequence of resting (CTRW) and ballistic (LW) phases continues indefinitely, defining a two-state random walk process.
The coexistence of localized trapping and long-range ballistic transport gives rise to rich anomalous diffusion behavior, which we analyze in the following sections.

\section{Constitutive Exponents of Anomalous Diffusion in TSRW}
\label{sec:tsrw_constitutive}

Diffusion in the Two-State Random Walk (TSRW) is characterized by the mean squared displacement (MSD),
\begin{equation}
\label{MSD_TSRW}
\langle x^2(t) \rangle \equiv \langle [x(t)-x(0)]^2 \rangle,
\end{equation}
whose long-time scaling
\begin{equation}
\langle x^2(t) \rangle \sim t^{2H}
\end{equation}
defines the \textit{Hurst exponent} $H$~\cite{alexander1969comments,mandelbrot2002gaussian}.
For processes obeying the Central Limit Theorem (CLT), $H=\tfrac12$, corresponding to normal diffusion.
When $H\neq\tfrac12$, at least one assumption of the CLT is violated and the process exhibits anomalous diffusion.

As discussed in the Introduction, deviations from normal diffusion can be traced to three distinct violations of the Central Limit Theorem (CLT): long-range temporal correlations between increments (Joseph effect), heavy-tailed increment statistics (Noah effect), and temporal nonstationarity or aging (Moses effect). These mechanisms are quantified by the scaling exponents $J$, $L$, and $M$, respectively, which together determine the Hurst exponent through the relation $H=J+L+M-1$.

To characterize these effects in the TSRW model, we introduce the statistical observables from which the exponents $J$, $L$, and $M$ are extracted. We then derive their asymptotic scaling behavior and identify the dominant constitutive mechanisms responsible for the observed diffusion regimes.

\subsection{Joseph Exponent $J$}
\label{subsec:tsrw_joseph}

The Joseph effect measures the contribution of long-range temporal correlations to anomalous diffusion. To quantify this effect, we examine the scaling behavior of the time-averaged mean-squared displacement (TAMSD),
\begin{equation}
\label{TAMSD_TSRW}
\Big\langle \overline{x^2(t,\Delta)} \Big\rangle
=
\left\langle
\frac{1}{t-\Delta}\int_{0}^{t-\Delta}
\big[x(t_0+\Delta)-x(t_0)\big]^2\,\mathrm{d}t_0
\right\rangle ,
\end{equation}
which exhibits the asymptotic scaling form
\begin{equation}
\label{J_scaling_TSRW}
\Big\langle \overline{x^2(t,\Delta)} \Big\rangle
\sim t^{\,2L+2M-2}\,\Delta^{2J}.
\end{equation}
The Joseph exponent satisfies $0\le J\le1$, where $J=\tfrac12$ corresponds to uncorrelated increments,
$J<\tfrac12$ indicates anti-correlations, and $J>\tfrac12$ signals persistent correlations.
In the TSRW, such correlations arise naturally from the intermittent switching between
localized CTRW phases and ballistic Lévy-walk phases.

\subsection{Moses Exponent $M$}
\label{subsec:tsrw_moses}

The Moses effect captures temporal non-stationarity and aging of the increments.
For continuous-time stochastic processes, it is quantified through the scaling of the first moment
of the absolute velocity~\cite{aghion2021moses},
\begin{equation}
\label{Moses_TSRW}
\langle |\mathrm{v}| \rangle \sim t^{\,M-\tfrac12}.
\end{equation}
In the TSRW, non-stationarity emerges from the stochastic alternation between resting (CTRW)
and moving (LW) phases, each characterized by distinct dynamical rules.
As a result, the Moses exponent need not vanish, even though the LW component alone
exhibits stationary increments.

\subsection{Noah Exponent $L$}
\label{subsec:tsrw_noah}

The Noah effect characterizes heavy-tailed increment statistics and is obtained from
the scaling of the second moment of the velocity~\cite{aghion2021moses},
\begin{equation}
\label{Noah_TSRW}
\langle \mathrm{v}^2 \rangle \sim t^{\,2L+2M-2}.
\end{equation}
The Noah exponent satisfies $L\ge\tfrac12$, with $L>\tfrac12$ indicating the dominance
of rare but large fluctuations.
In contrast to the standard Lévy walk, where the velocity is bounded and $L=\tfrac12$,
the TSRW can exhibit nontrivial Noah effects due to the stochastic coupling
between Gaussian jumps in the CTRW state and ballistic excursions in the LW state.

\subsection{Scaling Relation and Its Validity for TSRW}
\label{subsec:tsrw_scaling_relation}

The complete decomposition of anomalous diffusion was originally established in~\cite{chen2017anomalous}
and subsequently extended to continuous-time stochastic processes using the scaling Green--Kubo
formalism~\cite{meyer2017scale,aghion2021moses}. Under general conditions, this framework yields the
universal scaling relation
\begin{equation}
\label{scaling_relation_TSRW}
H = J + L + M - 1 .
\end{equation}

We now demonstrate that this relation remains valid for the Two-State Random Walk (TSRW).
In the TSRW model, the dynamics alternate between two statistically independent states:
a Lévy-walk (LW) state with nonzero velocity $\mathrm{v}_{\mathrm{LW}}(t)$ and a
continuous-time random walk (CTRW) state characterized by complete rest,
$\mathrm{v}_{\mathrm{CTRW}}(t)=0$.

The velocity autocorrelation function (VCF) can therefore be decomposed as
\begin{equation}
\begin{aligned}
\langle \mathrm{v}(t)\mathrm{v}(t+\Delta)\rangle
=&\;\langle \mathrm{v}_{\mathrm{LW}}(t)\mathrm{v}_{\mathrm{LW}}(t+\Delta)\rangle\\
&+ \langle \mathrm{v}_{\mathrm{LW}}(t)\mathrm{v}_{\mathrm{CTRW}}(t+\Delta)\rangle \\
&\;+\langle \mathrm{v}_{\mathrm{CTRW}}(t)\mathrm{v}_{\mathrm{LW}}(t+\Delta)\rangle\\
&+ \langle \mathrm{v}_{\mathrm{CTRW}}(t)\mathrm{v}_{\mathrm{CTRW}}(t+\Delta)\rangle .
\end{aligned}
\end{equation}

Since the CTRW state has zero velocity at all times and is statistically independent of the LW state,
all mixed and CTRW-only contributions vanish identically,
\begin{align}
\langle \mathrm{v}_{\mathrm{LW}}(t)\mathrm{v}_{\mathrm{CTRW}}(t+\Delta)\rangle &= 0, \\
\langle \mathrm{v}_{\mathrm{CTRW}}(t)\mathrm{v}_{\mathrm{LW}}(t+\Delta)\rangle &= 0, \\
\langle \mathrm{v}_{\mathrm{CTRW}}(t)\mathrm{v}_{\mathrm{CTRW}}(t+\Delta)\rangle &= 0.
\end{align}
As a result, the VCF reduces to
\begin{equation}
\label{vcf_tsrw}
\langle \mathrm{v}(t)\mathrm{v}(t+\Delta)\rangle
=
\langle \mathrm{v}_{\mathrm{LW}}(t)\mathrm{v}_{\mathrm{LW}}(t+\Delta)\rangle .
\end{equation}

This shows that velocity correlations in the TSRW are entirely determined by the Lévy-walk
segments, despite the intermittent interruptions by CTRW resting phases.
Consequently, the scaling Green--Kubo formalism remains applicable, and the scaling relation
\eqref{scaling_relation_TSRW} holds for the TSRW as well~\cite{aghion2021moses}.

In the following sections, we compute the Noah and Moses exponents $L$ and $M$ from the velocity
statistics of the TSRW, while the Joseph exponent $J$ is extracted from the dominant scaling of the
time-averaged mean squared displacement (TAMSD). The Hurst exponent $H$ is obtained independently
from the asymptotic scaling of the MSD. Together, these results enable a complete constitutive
decomposition of anomalous diffusion in the TSRW and reveal how the stochastic alternation between
CTRW and Lévy-walk phases generates nontrivial Joseph, Noah, and Moses effects.

\section{Hurst Exponent for the TSRW Model}

The Hurst exponent $H$ of the Two-State Random Walk (TSRW) can be determined from the long-time scaling of the mean squared displacement (MSD). The MSD for the TSRW was derived analytically in Ref.~\cite{liu2022strong}, and we briefly summarize the essential results here.

In the TSRW model, the probability density of finding a particle at position $x$ at time $t$ is composed of two contributions,
\begin{equation}
P(x,t) = P_r(x,t) + P_j(x,t),
\end{equation}
where $P_r(x,t)$ corresponds to particles in the resting (CTRW) phase and $P_j(x,t)$ to particles in the walking (LW) phase.

For an unbiased process, the MSD is obtained from the second moment of the total propagator,
\begin{equation}
\langle x^2(t) \rangle
= - \mathscr{L}^{-1}\!\left[
\left.\frac{\partial^2 P(k,s)}{\partial k^2}\right|_{k=0}
\right],
\end{equation}
where $P(k,s)$ is the Fourier–Laplace transform of $P(x,t)$ and $\mathscr{L}^{-1}$ denotes the inverse Laplace transform.

According to Ref.~\cite{liu2022strong}, the Fourier–Laplace transforms of the resting and walking contributions are given by
\begin{equation}
P_r(k,s) =
\frac{\Phi_r(s)\,P_0(k)}
{1 - \tfrac{1}{2}\!\left[\omega_j(s+ick)+\omega_j(s-ick)\right]\lambda(k)\omega_r(s)},
\end{equation}
\begin{equation}
P_j(k,s) =
\frac{\omega_r(s)\,\Phi_j(k,s)\,P_0(k)}
{1 - \tfrac{1}{2}\!\left[\omega_j(s+ick)+\omega_j(s-ick)\right]\lambda(k)\omega_r(s)},
\end{equation}
so that
\begin{equation}
P(k,s)
= \frac{\Phi_r(s)+\omega_r(s)\Phi_j(k,s)}
{1 - \tfrac{1}{2}\!\left[\omega_j(s+ick)+\omega_j(s-ick)\right]\lambda(k)\omega_r(s)}.
\end{equation}

Here, $\Phi_r(t)=1-\int_0^t\omega_r(t')dt'$ is the survival probability of the CTRW phase, and
\begin{equation}
\Phi_j(x,t)=\tfrac12\,\delta(|x|-vt)\,\Phi_j(t),
\qquad
\Phi_j(t)=1-\int_0^t\omega_j(t')dt',
\end{equation}
is the survival probability of the LW phase. Using these expressions, the MSD can be computed analytically.

The resulting asymptotic scaling of the MSD depends on the exponents $\alpha$ and $\beta$ of the waiting-time distributions $\omega_r(\tau)\sim\tau^{-1-\alpha}$ and $\omega_j(\tau)\sim\tau^{-1-\beta}$. The dominant long-time contributions are summarized in Table~I.

\begin{table}[t]
\centering
\caption{Asymptotic scaling of the mean squared displacement (MSD) for the TSRW model in different parameter regimes. The dominant long-time contribution determines the Hurst exponent $H$.}
\label{table:MSD_TSRW}
\begin{tabular}{ccc}
\hline
Regime & Parameter Range & MSD Scaling \\ \hline
I   & $0<\beta<\alpha<1$ 
    & $\langle x^2(t)\rangle \sim t^{2}$ \\[4pt]

II  & $0<\alpha<\beta<1$ 
    & $\langle x^2(t)\rangle \sim t^{2+\alpha-\beta}$ \\[4pt]

III & $1<\alpha<2,\; 1<\beta<2$ 
    & $\langle x^2(t)\rangle \sim t^{3-\beta}$ \\[4pt]

IV  & $1<\alpha<2,\; 0<\beta<1$ 
    & $\langle x^2(t)\rangle \sim t^{2}$ \\[4pt]

V   & $0<\alpha<1,\; 1<\beta<2$ 
    & $\langle x^2(t)\rangle \sim t^{2+\alpha-\beta}$ \\ \hline
\end{tabular}
\end{table}

To extract the Hurst exponent $H$, we retain only the leading term in the long-time limit, since the contribution with the largest exponent dominates the MSD scaling. The Hurst exponent is then defined through
\begin{equation}
\langle x^2(t) \rangle \sim t^{2H},
\end{equation}
with $H$ determined by the largest power of $t$ in the corresponding parameter regime.




\subsection{J for VGLW}
For TSRW, we can also determine the J exponent from EATAMSD. For TSRW, it is divided into two things

1. EATAMSD for CTRW
2. EATAMSD for LW

which has been calculated in  \cite{liu2022coexistence}.  EATAMSD for CTRW has been calculated using renewal events, whereas EATAMSD for LW has been calculated using the velocity correlation function  \cite{liu2022coexistence, meyer2017scale}. 
The EATAMSD of TSRW is shown in the Table. \ref{Specific values of the lag-time exponents of the TSRW model.}, calculated \cite{liu2022coexistence}, 
\begin{table}[tbh]
\centering
\begin{tabular}{|c|c|lll}
\cline{1-2}
\textbf{Range of Parameters}        & $\boldsymbol{\left\langle\left\langle x^2(\tau) \right\rangle_T\right\rangle_E}$   &  &  &  \\ \cline{1-2}
$0<\beta<1$     & $\sim \tau^2+\sim \tau^1$                 &  &  &  \\ \cline{1-2}
$1<\beta<2$     & $\sim \tau^{3-\beta}+\sim \tau^1$ &  &  &  \\ \cline{1-2}
\end{tabular}
\caption{Specific values of the lag-time exponents of the TSRW model.}
\label{Specific values of the lag-time exponents of the TSRW model.}
\end{table}

\begin{figure}[tbh]	
\centering
\includegraphics[width=0.49\linewidth]{"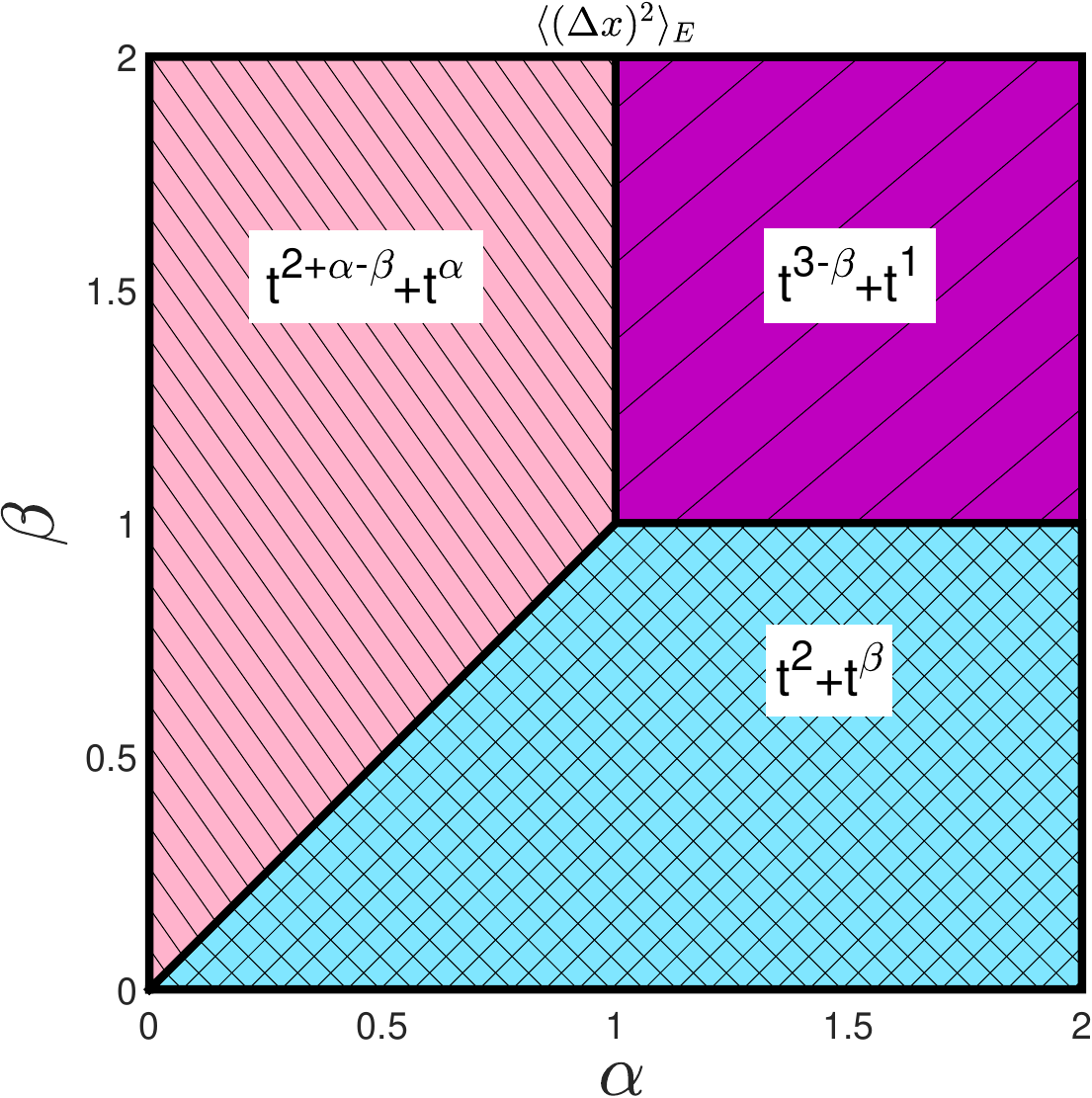"}
\includegraphics[width=0.49\linewidth]{"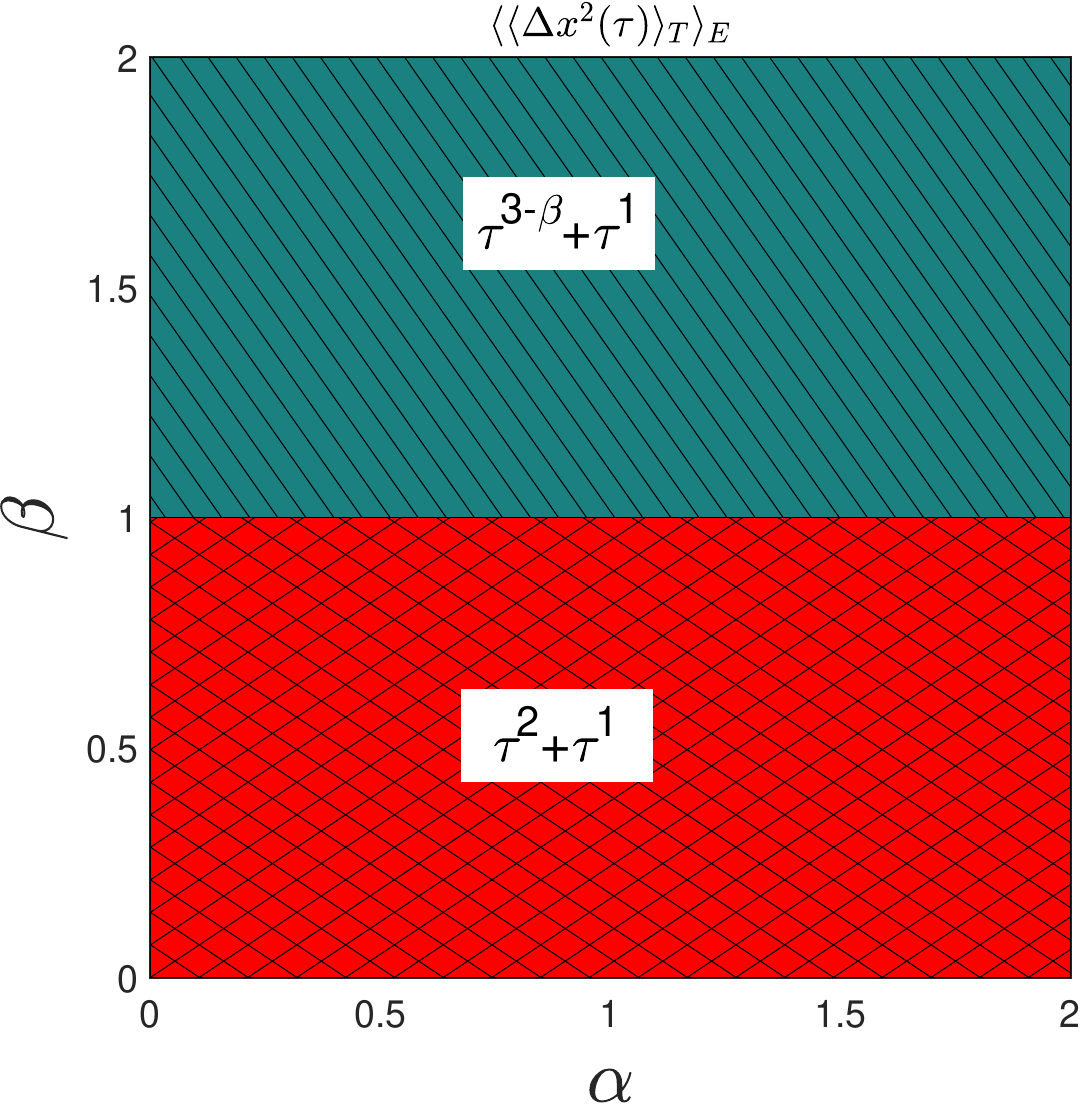"}\\{\footnotesize (a)}\hspace{1.5in} {\footnotesize (b)}
\caption{(a) $\left\langle\Delta \mathbf{x}^2(\tau)\right\rangle_{\mathrm{E}}$(b)$\left\langle\left\langle\Delta \mathbf{x}^2(\tau)\right\rangle_{\mathrm{T}}\right\rangle_{\mathrm{E}}$ of two state random walk model.}
\label{easd_eatasd_tsrw}
\end{figure}

\begin{figure}[tbh]	
\centering
\includegraphics[width=0.49\linewidth]{"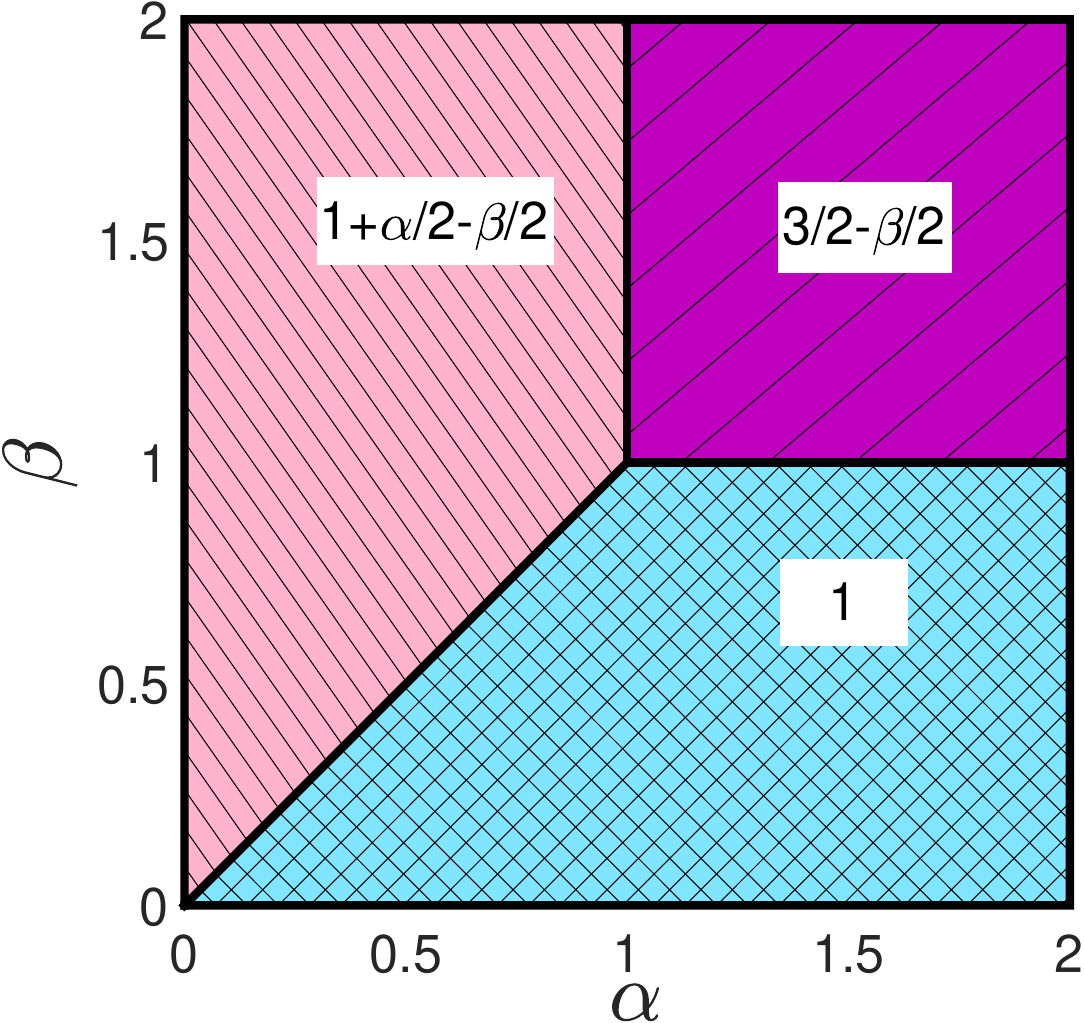"}
\includegraphics[width=0.49\linewidth]{"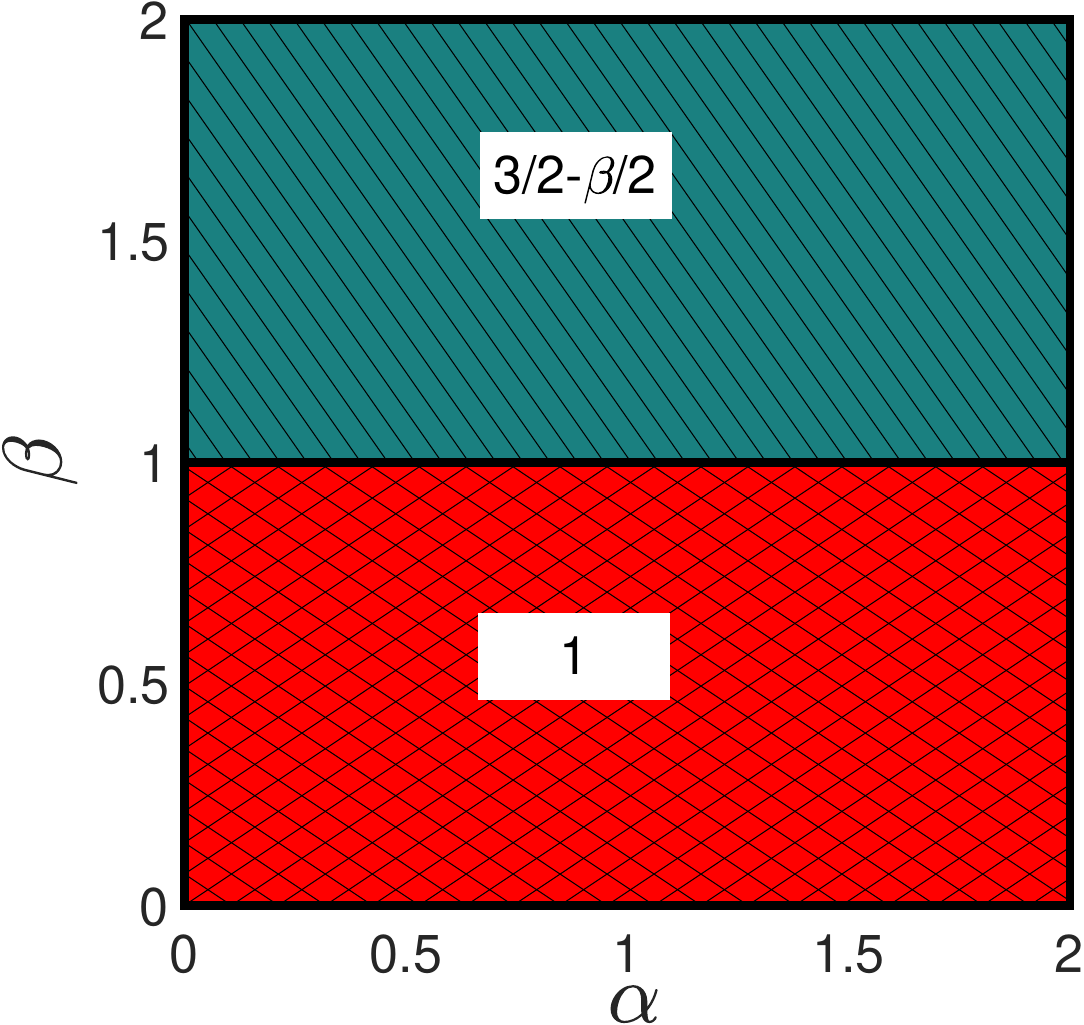"}\\{\footnotesize (a)}\hspace{1.5in} {\footnotesize (b)}
	\caption{The phase portrait of (a) $H$ and (b) $J$ effect in two state random walk model.}
\label{Hurst_Joseph_two_state}
 \end{figure}

To determine the Joseph exponent, we have chosen the lag-time with the largest exponent from the EATAMSD as the component with the largest exponent will determine the behaviour of $\left\langle\left\langle x^2(\tau) \right\rangle_T\right\rangle_E$ in the long time limit $T\rightarrow\infty$.

\begin{equation}
J= \begin{cases}1, & 0<\beta<1 \\ (3-\beta) / 2, & 1<\beta<2 \end{cases}
\end{equation}
Fig. \ref{easd_eatasd_tsrw}.a and fig. \ref{easd_eatasd_tsrw}.b show the phase portrait of TSRW summarizing different regimes based on the EASD and EATAMSD.
Fig. \ref{Hurst_Joseph_two_state}.a and fig. \ref{Hurst_Joseph_two_state}.b show the phase portrait of TSRW summarizing different regimes based on the Hurst and Joseph exponents.
\section{Velocity Propagator of the TSRW}

To compute the velocity propagator of the Two-State Random Walk (TSRW), denoted by
$p(\mathrm{v},t)$, we exploit the fact that the dynamics alternate between two distinct
states:
(i) a CTRW (resting) phase in which the particle is immobile and hence has zero velocity,
and
(ii) a Lévy-walk (LW) phase in which the particle moves ballistically with constant speed
$c$ in either direction.

As a result, the velocity propagator is given by
\begin{equation}
\label{eqn:Velocity_propagator}
p(\mathrm{v},t)
=
P_j(t)\,\frac{1}{2}\big[\delta(\mathrm{v}-c)+\delta(\mathrm{v}+c)\big]
+
P_r(t)\,\delta(\mathrm{v}),
\end{equation}
where $P_j(t)$ and $P_r(t)$ denote the probabilities that the particle is in the LW and
CTRW phases, respectively, at time $t$.

Following Ref.~\cite{liu2022coexistence}, these probabilities can be expressed in Laplace
space as
\begin{align}
P_r(t) &= \mathcal{L}^{-1}\!\left[\frac{\Phi_r(s)}{1-\omega_r(s)\omega_j(s)}\right],\\
P_j(t) &= \mathcal{L}^{-1}\!\left[\frac{\omega_r(s)\Phi_j(s)}{1-\omega_r(s)\omega_j(s)}\right],
\end{align}
where $\omega_r(s)$ and $\omega_j(s)$ are the Laplace transforms of the resting and jumping
time distributions, and $\Phi_r(s)$ and $\Phi_j(s)$ are the corresponding survival
probabilities.

\subsection{$0<\alpha<\beta<1$}

In this regime, the Laplace transform of $P_j(t)$ behaves as
\begin{equation}
P_j(s)\simeq \frac{\tau_\beta}{\tau_\alpha}\, s^{-(1+\alpha-\beta)},
\end{equation}
which yields, upon inversion,
\begin{equation}
P_j(t)=\frac{\tau_\beta}{\Gamma(1+\alpha-\beta)\,\tau_\alpha}\, t^{\alpha-\beta}.
\end{equation}

\subsection{$0<\alpha<1<\beta<2$}

Here the dominant contribution is
\begin{equation}
P_j(s)\simeq \frac{T_\beta}{\tau_\alpha}\, s^{-\alpha},
\end{equation}
leading to
\begin{equation}
P_j(t)=\frac{T_\beta}{\Gamma(\alpha)\,\tau_\alpha}\, t^{\alpha-1}.
\end{equation}

\subsection{$0<\beta<\alpha<1$ and $0<\beta<1<\alpha<2$}

In both cases one finds
\begin{equation}
P_j(s)\simeq \frac{1}{s},
\end{equation}
implying that
\begin{equation}
P_j(t)=\text{const}.
\end{equation}

\subsection{$1<\min(\alpha,\beta)<2$}

When both exponents exceed unity, the Laplace transform reduces to
\begin{equation}
P_j(s)\simeq \frac{T_\beta}{(T_\alpha+T_\beta)s},
\end{equation}
and hence
\begin{equation}
P_j(t)=\frac{T_\beta}{T_\alpha+T_\beta},
\end{equation}
which is time independent.





\section{$L$ and $M$ Exponents for the TSRW}

The Noah and Moses exponents can be obtained directly from the first and second moments
of the velocity propagator in Eq.~\eqref{eqn:Velocity_propagator}.

\subsection{$0<\alpha<\beta<1$}

In this regime, both moments scale as
$\langle|\mathrm{v}|\rangle \sim \langle \mathrm{v}^2\rangle \sim t^{\alpha-\beta}$,
yielding
\begin{equation}
M=\alpha-\beta+\tfrac{1}{2},
\qquad
L=\tfrac{1}{2}+\tfrac{\beta-\alpha}{2}.
\end{equation}

\subsection{$0<\alpha<1<\beta<2$}

Here the velocity moments scale as
$\langle|\mathrm{v}|\rangle \sim \langle \mathrm{v}^2\rangle \sim t^{\alpha-1}$,
which gives
\begin{equation}
M=\alpha-\tfrac{1}{2},
\qquad
L=1-\tfrac{\alpha}{2}.
\end{equation}

\subsection{$0<\beta<\alpha<1$, $0<\beta<1<\alpha<2$, and $1<\min(\alpha,\beta)<2$}

In these regimes both velocity moments remain finite,
$\langle|\mathrm{v}|\rangle \sim \langle \mathrm{v}^2\rangle \sim t^0$,
leading to
\begin{equation}
M=\tfrac{1}{2},
\qquad
L=\tfrac{1}{2}.
\end{equation}

\begin{figure}[t]
\centering
\includegraphics[width=0.48\linewidth]{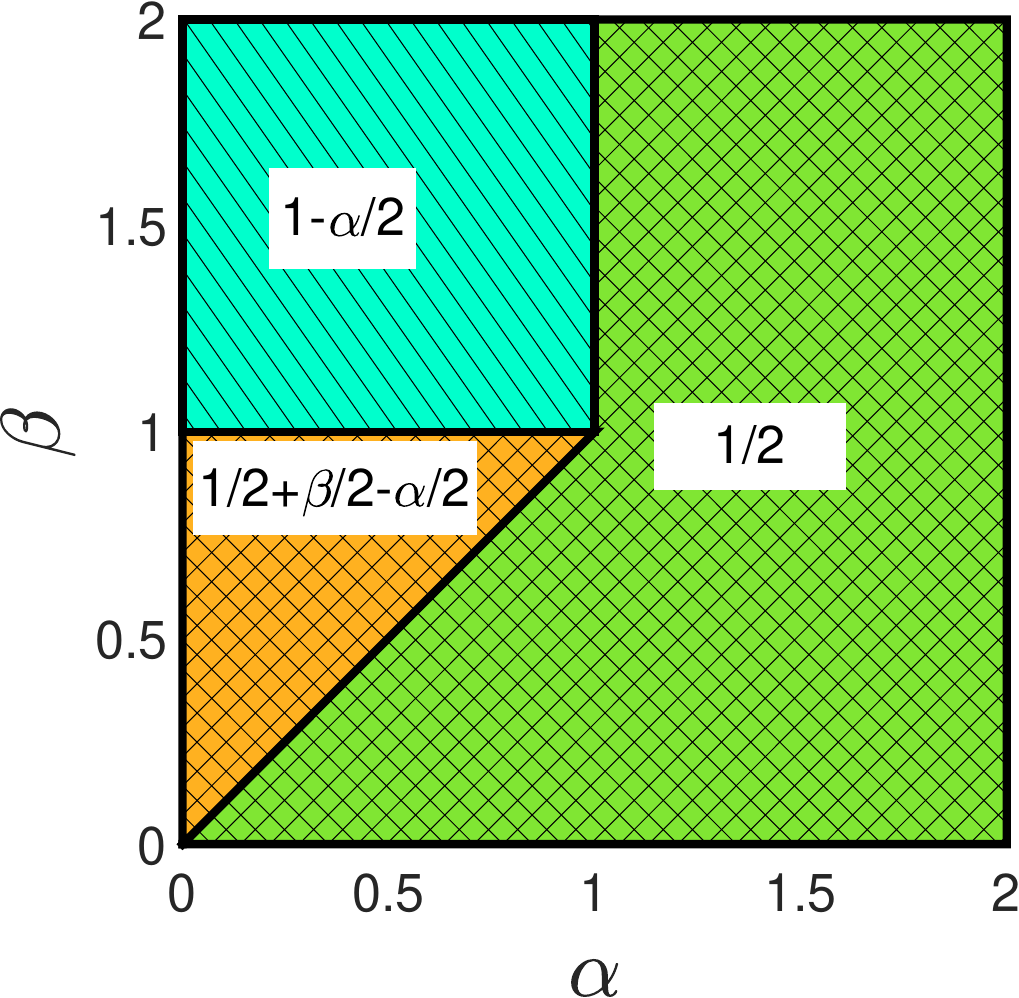}
\includegraphics[width=0.48\linewidth]{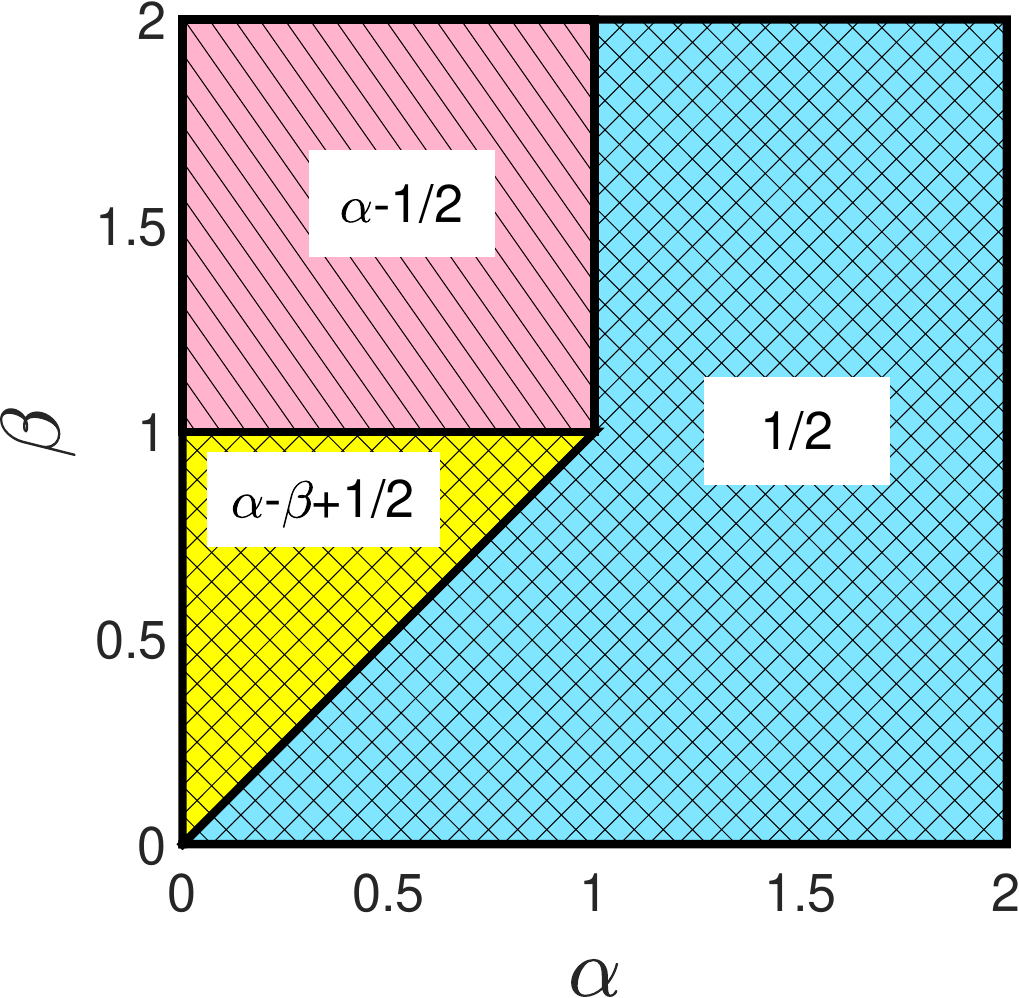}
\\{\footnotesize (a)}\hspace{1.5in} {\footnotesize (b)}
\caption{Phase portraits of (a) the Noah exponent $L$ and (b) the Moses exponent $M$
for the Two-State Random Walk.}
\label{fig:LM_TSRW}
\end{figure}

\section{Phase Decomposition of the TSRW Model}

We now summarize the full phase decomposition of the Two-State Random Walk (TSRW) model
in terms of the Hurst exponent $H$ and the constitutive exponents
$J$ (Joseph), $L$ (Noah), and $M$ (Moses).
The phase space is partitioned into four distinct regions,
determined by the relative values of the power-law exponents
$\alpha$ (CTRW waiting times) and $\beta$ (LW flight times).
Each region is characterized by a different combination of the three
mechanisms responsible for anomalous diffusion.

\begin{figure}[t]
    \centering
    \includegraphics[width=0.7\columnwidth]{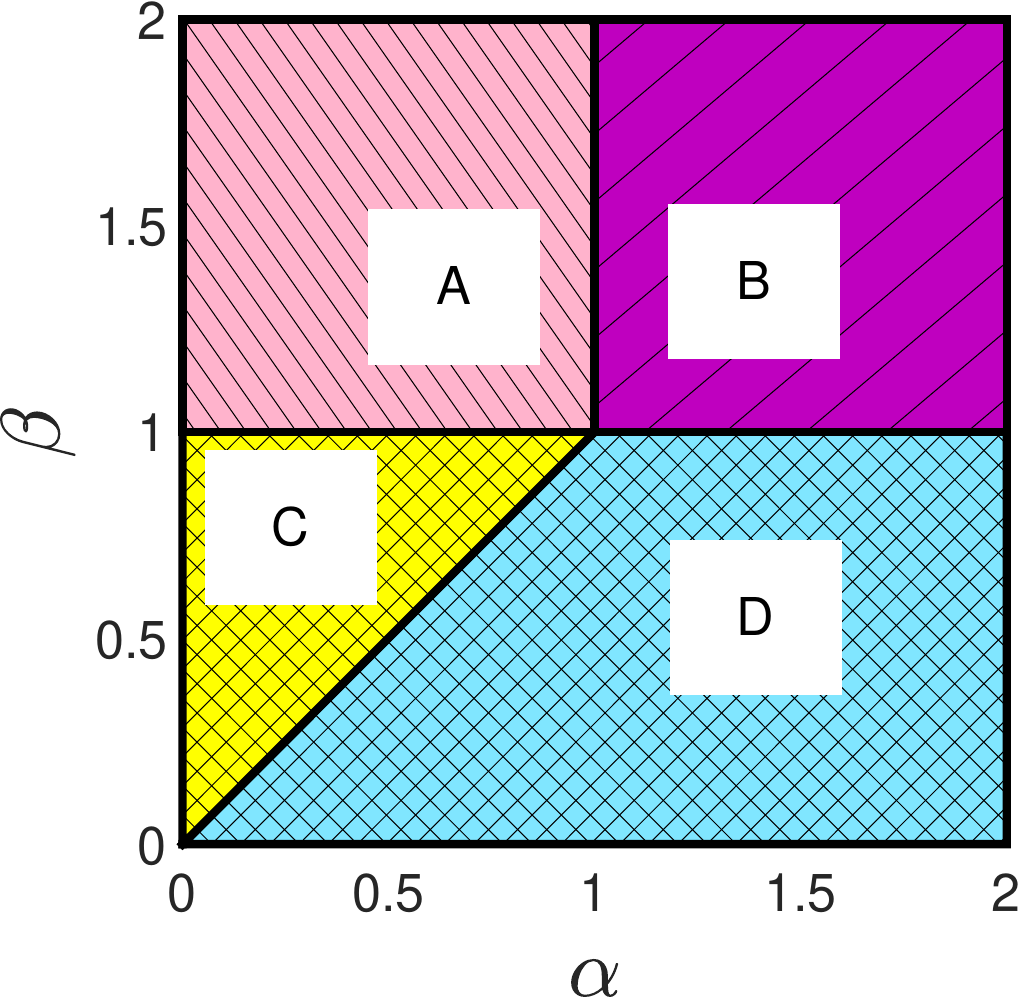}
    \caption{Overall phase diagram of the Two-State Random Walk, showing the regions
    classified by the scaling exponents $H$, $J$, $L$, and $M$.}
    \label{fig:Overall_phase_TSRW}
\end{figure}

\begin{enumerate}

\item \textbf{Region A: $0 < \alpha < 1 < \beta < 2$}

The scaling exponents are
\[
H = \tfrac{1}{2} + \tfrac{\alpha}{2} - \tfrac{\beta}{2}, \quad
J = \tfrac{3}{2} - \tfrac{\beta}{2}, \quad
L = 1 - \tfrac{\alpha}{2}, \quad
M = \alpha - \tfrac{1}{2}.
\]
In this regime, anomalous diffusion is generated by the combined action of
all three constitutive effects.
The Joseph effect is present but not maximal,
while the long trapping times in the CTRW phase lead to pronounced
non-stationarity (Moses effect) and heavy-tailed increment statistics (Noah effect).

\item \textbf{Region B: $1 < \min(\alpha,\beta) < 2$}

The scaling exponents reduce to
\[
H = \tfrac{3}{2} - \tfrac{\beta}{2}, \quad
J = \tfrac{3}{2} - \tfrac{\beta}{2}, \quad
L = \tfrac{1}{2}, \quad
M = \tfrac{1}{2}.
\]
Here, anomalous diffusion arises solely from the Joseph effect.
Both the Noah and Moses effects are absent, indicating that the increment
distribution is stationary and possesses finite moments.
Correlations persist, but they are not maximal.

\item \textbf{Region C: $0 < \alpha < \beta < 1$}

The scaling exponents are
\[
H = 1 + \tfrac{\alpha}{2} - \tfrac{\beta}{2}, \quad
J = 1, \quad
L = \tfrac{1}{2} + \tfrac{\beta}{2} - \tfrac{\alpha}{2}, \quad
M = \alpha - \beta + \tfrac{1}{2}.
\]
In this region, all three effects contribute to anomalous diffusion.
The Joseph effect is maximal, reflecting strong temporal correlations,
while the Noah and Moses effects originate from the heavy-tailed trapping
times in the CTRW state.

\item \textbf{Region D: $0 < \beta < 1$ and $\beta < \alpha < 2$}

The scaling exponents take the universal values
\[
H = 1, \quad
J = 1, \quad
L = \tfrac{1}{2}, \quad
M = \tfrac{1}{2}.
\]
Anomalous diffusion in this regime is entirely driven by a maximal Joseph effect.
Both the Noah and Moses effects are absent, indicating stationary increments
with finite variance.

\end{enumerate}

Figure~\ref{fig:Overall_phase_TSRW} provides a global overview of the TSRW phase diagram,
highlighting how the interplay between the CTRW and LW components generates
distinct anomalous diffusion mechanisms across parameter space.

\section{Numerical Simulations}
\begin{figure}[tbh]
	\centering
	\includegraphics[width=0.75\columnwidth]{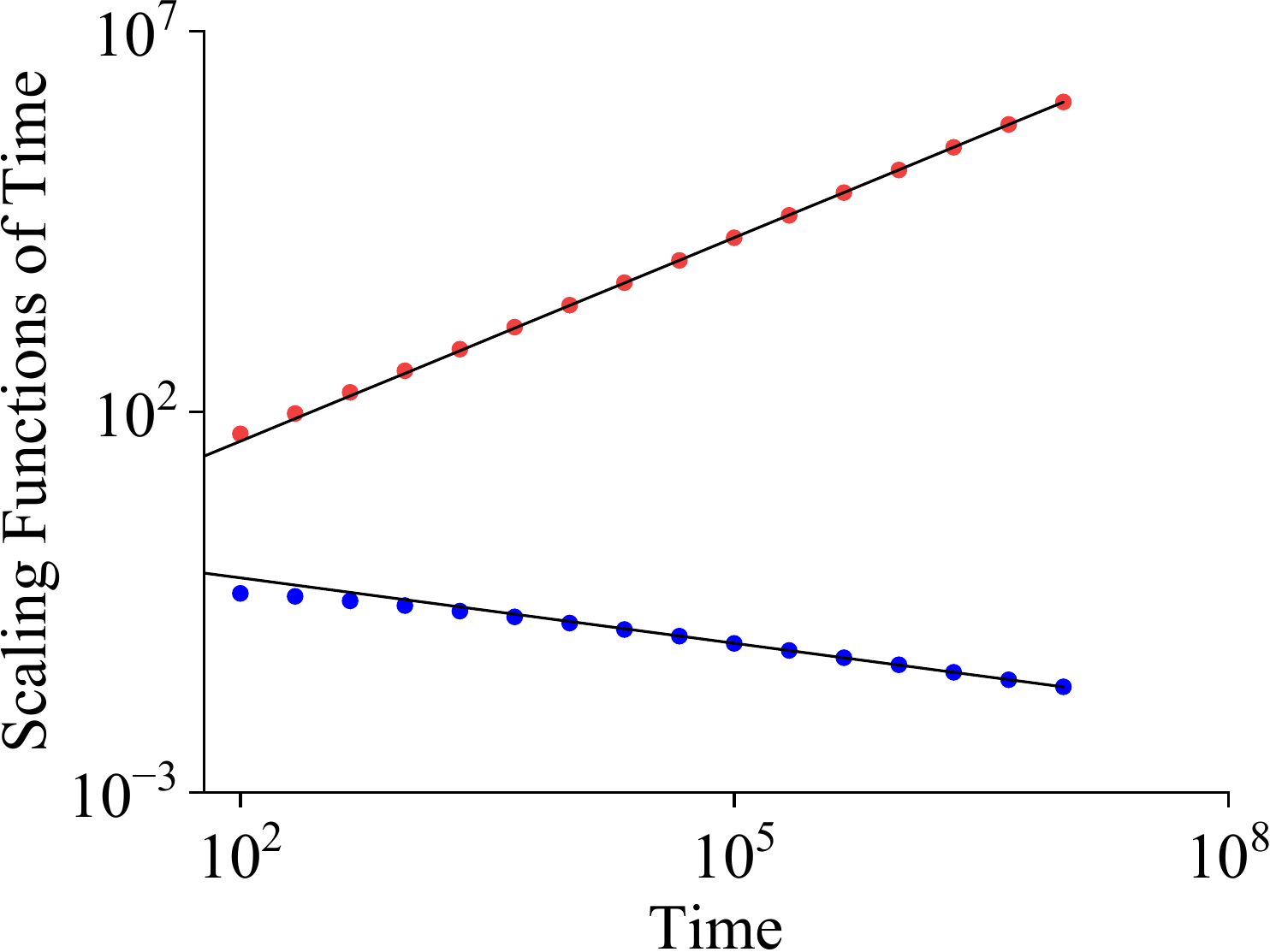}\\
	{\footnotesize (a)}\\
	\includegraphics[width=0.75\columnwidth]{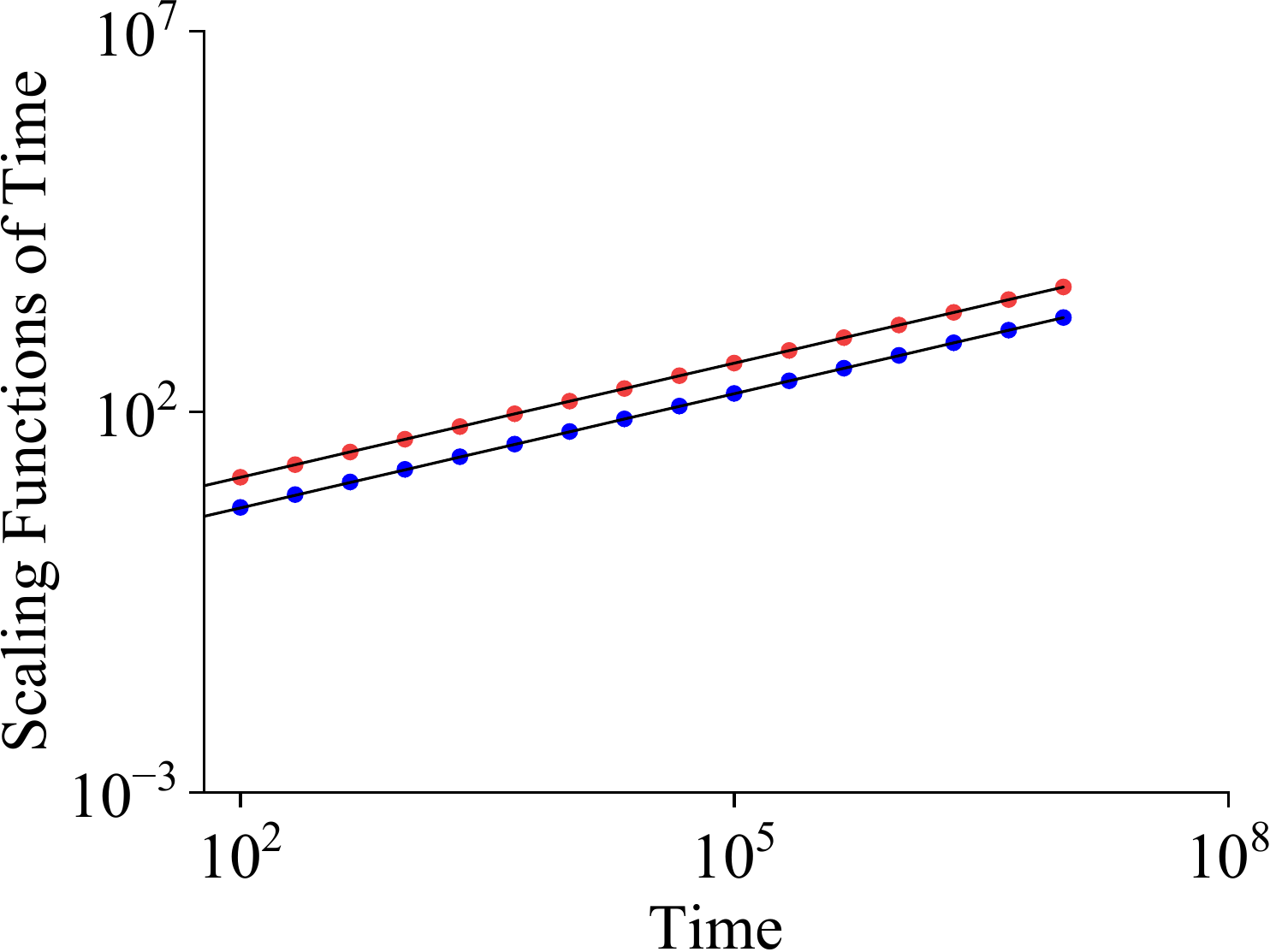}\\
	{\footnotesize (b)}\\
	\includegraphics[width=0.75\columnwidth]{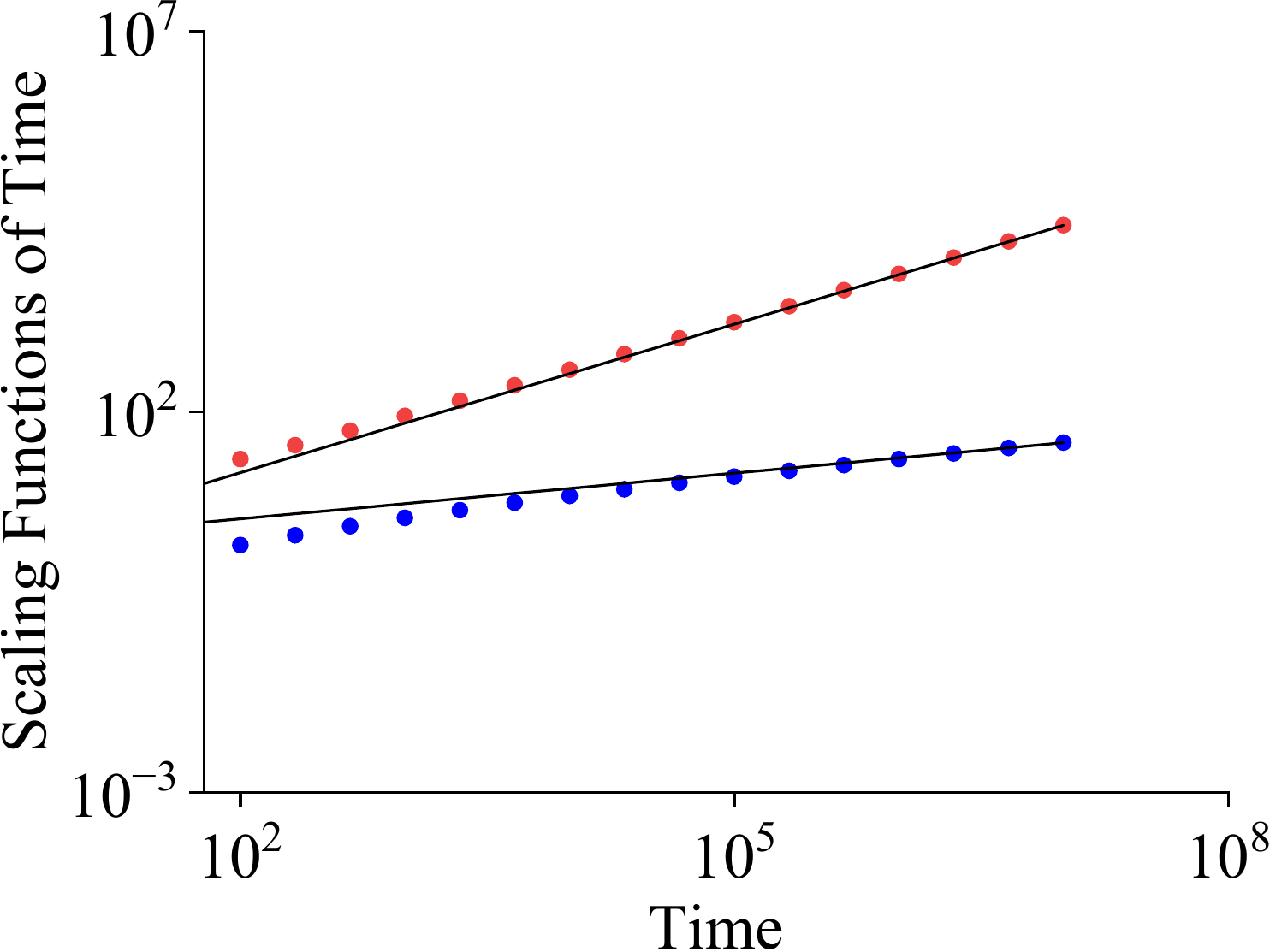}\\
	{\footnotesize (c)}\\
	\includegraphics[width=0.75\columnwidth]{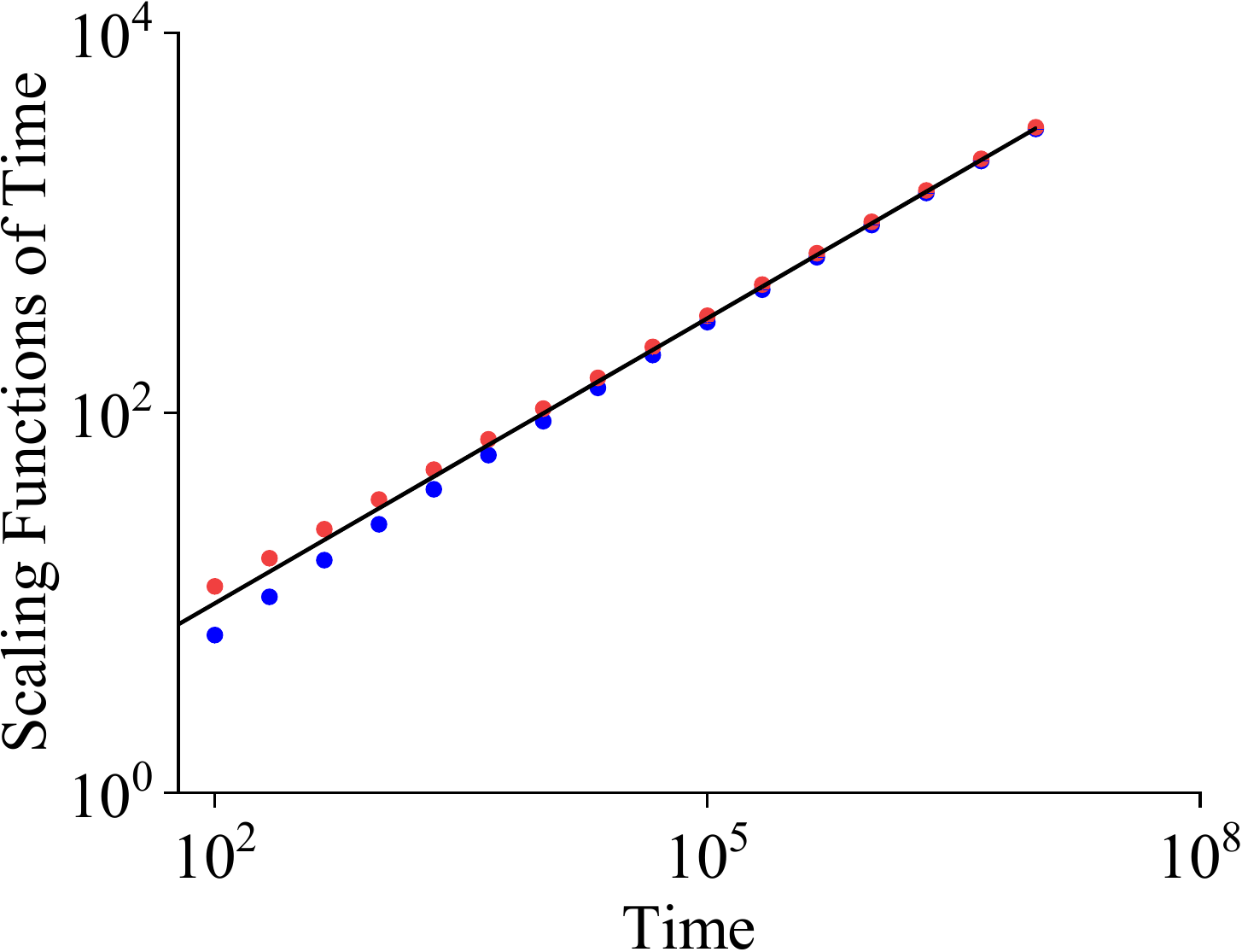}\\
	{\footnotesize (d)}
	\caption{$t^M$ and $t^L$ upto
	$t=10^7$, obtained from ensemble average of $10^6$ realizations.
    (a) $\alpha=0.2$ and $\beta=1.5$;
	(b) $\alpha=1.3$ and $\beta=1.3$;
	(c) $\alpha=0.5$ and $\beta=0.2$;
	(d) $\alpha=0.2$ and $\beta=0.5$.
    The points represent simulation results, while the black solid lines denote the fitted lines. The blue and red dots denote the $t^M$ and $t^L$, respectively.}
	\label{L_M_tsrw}
\end{figure}

To confirm our analytical predictions, we numerically computed the $L$ and $M$ exponents using an ensemble average over $10^6$ random walkers. Figure~\ref{L_M_tsrw} shows the numerically calculated scaling functions, $t^M$ and $t^L$, as blue and red dots, respectively, while the black solid lines denote the analytical predictions.
\section{Discussion and Conclusion}
The Two-State Random Walk (TSRW) provides a minimal yet powerful framework for describing
transport processes that alternate between localized trapping and ballistic motion.
By construction, the model interpolates between two fundamental stochastic dynamics:
a Continuous-Time Random Walk (CTRW) phase, characterized by random waiting times and
zero velocity, and a Lévy Walk (LW) phase, characterized by persistent ballistic motion
with finite speed. Through the interplay of these two states, the TSRW captures a broad
class of intermittent transport phenomena that cannot be described by either mechanism
alone.

In Refs.~\cite{PhysRevLett.71.3975,WEEKS1996291,SOLOMON199470}, the transport of tracer particles in a two-dimensional rotating annulus was described using a similar two-state model. In that model, tracer particles alternate between being trapped near vortices and having motion with constant velocity through jet regions. The duration of both types of motion have a power-law distribution. If the CTRW state in the TSRW model is constrained to allow complete rest rather than displacement, the resulting dynamics would be the same.

In this paper, we have analyzed anomalous diffusion in the TSRW by decomposing it into its three
constitutive mechanisms—the Joseph, Noah, and Moses effects. This decomposition reveals
four distinct dynamical regimes in the TSRW phase space, each governed by a different
combination of temporal correlations, heavy-tailed fluctuations, and aging effects.
Although the LW component alone exhibits only Joseph-type correlations and is devoid of
both Noah and Moses effects due to its fixed-speed and stationary increment structure,
the incorporation of the CTRW phase fundamentally alters the statistical properties of
the process. In particular, when the CTRW waiting-time distribution has a divergent mean,
the combined dynamics generically generate both heavy-tailed velocity fluctuations and
non-stationarity, activating the Noah and Moses effects in the full TSRW.

A key result of our analysis is that the scaling relation
\begin{equation}
H = J + L + M - 1
\end{equation}
holds throughout the entire TSRW phase space. Despite the hybrid and non-Markovian nature
of the dynamics, the Green--Kubo-based scaling framework remains valid, with the velocity
autocorrelation function determined solely by the LW segments. This demonstrates the
robustness of the anomalous diffusion decomposition when applied to composite stochastic
processes with dynamically switching states.

For parameter regimes in which both the CTRW and LW waiting-time distributions possess
finite means, the TSRW exhibits anomalous diffusion driven exclusively by the Joseph
effect, with $L = M = 1/2$. In contrast, when the CTRW phase is characterized by long
trapping times, the system enters regimes where all three constitutive effects contribute
simultaneously. In particular, strong aging in the CTRW phase produces a nontrivial Moses
effect, while intermittent switching between rest and ballistic motion generates
effective heavy-tailed velocity statistics, leading to Noah effect even though
neither component alone possesses such behavior.

These findings demonstrate that anomalous diffusion in the TSRW is not merely a
superposition of its constituent dynamics but rather an emergent phenomenon arising from
their interaction. The TSRW thus provides a clear example of how hybrid stochastic
processes can exhibit qualitatively new transport properties that are absent in their
individual components.

Beyond its theoretical significance, the TSRW framework is directly relevant to a wide
range of experimental systems exhibiting intermittent dynamics, including intracellular
transport with binding--unbinding events, animal movement with alternating search and
relocation phases, and transport in disordered or heterogeneous media. By identifying
the precise mechanisms responsible for anomalous scaling in each regime, our results
provide a systematic framework for interpreting single-particle tracking data and for
distinguishing between correlation-dominated, fluctuation-dominated, and aging-driven
transport.

Overall, this work establishes the TSRW as a prototypical hybrid model for anomalous
diffusion and demonstrates that the Joseph--Noah--Moses decomposition offers a unified
and physically transparent language for understanding anomalous transport in complex
systems with stochastic switching dynamics.

\bibliography{levy_walk}
\end{document}